\documentclass[aps,preprint,showpacs]{revtex4}

\usepackage{epsfig,color}
\usepackage{amsmath}

\newcommand{\pderiv}[2]{\frac{\partial #1}{\partial #2}}
\newcommand{\deriv}[2]{\frac{d #1}{d #2}}
\newcommand{\infint}{\int \limits_{-\infty}^{\infty}}
\definecolor{red}{rgb}{1,0,0}

\begin{document}

\title{A General Nonlinear Fokker-Planck Equation and its Associated Entropy}
\author{Veit Schw\"ammle, Evaldo M. F. Curado, Fernando D. Nobre}
\affiliation{Centro Brasileiro de Pesquisas F\'isicas, 
Rua Xavier Sigaud 150, Rio de Janeiro, RJ 22290-180, Brazil}

\date{\today}

\begin{abstract}
A recently introduced nonlinear Fokker-Planck equation, derived
directly from a master equation, comes out as a very 
general tool to describe phenomenologically systems presenting complex
behavior, like
anomalous diffusion, in the presence of  
external forces. Such an equation is characterized by a nonlinear diffusion
term that  
may present, in general, two distinct powers of the probability
distribution. Herein, we calculate the stationary-state distributions  
of this equation in some special cases, and introduce associated classes of
generalized entropies in order to 
satisfy the H-theorem. Within this approach, the parameters associated
with the 
transition rates of the original master-equation are related to such
generalized entropies, and are shown to obey some restrictions. Some
particular cases are discussed.  
 
 \vskip \baselineskip


\end{abstract}

\newpage

\maketitle

\section{Introduction}

The standard statistical-mechanics formalism, as proposed originally 
by Boltzmann and
Gibbs (BG), is considered as one of the most successful 
theories of physics, and it has enabled physicists to propose theoretical
models in order to derive
thermodynamical properties for real systems, by approaching the problem 
from the microscopic scale. Such a prescription has led to an
adequate description of a large diversity of physical systems, essentially
those represented by linear equations and characterized by short-range 
interactions and/or short-time memories.
Although BG statistical mechanics is well
formulated (under certain restrictions) for systems at equilibrium, 
the same is not true for
out-of-equilibrium systems, in such a way that most of this theory applies
only near equilibrium~\cite{Reichl:98,VanKampen:81,Haken:77}. One of the most
important phenomenological equations of nonequilibrium statistical
mechanics is the linear 
Fokker-Planck equation (FPE), that rules the time evolution of the
probability distribution associated with a given physical system, in the
presence of an external force field~\cite{Risken:89},  
provided that the states of the system can be expressed
by a continuum. This equation deals satisfactorily with many
physical situations, e.g., those associated with normal diffusion, and is
essentially associated with the BG formalism, in the sense that 
the Boltzmann distribution, which is
usually obtained through the maximization of the
BG entropy under 
certain constraints (the so-called MaxEnt principle), also appears as the 
stationary solution of the linear FPE~\cite{Risken:89,Cover:91}.

Nevertheless, restrictions to the applicability of the BG
statistical-mechanics formalism have been
found in many systems, including for instance, 
those characterized by nonlinearities, long-range interactions and/or
long-time memories, 
which may present several types of anomalous
behavior, e.g., stationary states far from 
equilibrium~\cite{next:99,next:04,next:05}.
These anomalous behaviors suggest that a more general
theory is required;
as a consequence of that, many attempts have been made, 
essentially by proposing generalizations
of the BG entropy
\cite{Tsallis:88,Abe:97,Anteneodo:99,Borges:98,Landsberg:98,%
Curado:99,CuradoNobre:2004,Renyi:70,Kaniadakis:01,Kaniadakis:02}. 
Among these, 
the most successful proposal, so far, appears to be the one suggested by
Tsallis~\cite{Tsallis:88}, through the introduction of a generalized
entropy, characterized by an index $q$, in such a way that the BG entropy 
is recovered in the limit 
$q \rightarrow 1$. The usual extensivity property of some thermodynamic
quantities holds only for $q=1$, and if $q \neq 1$ such
quantities do not increase linearly with the size of the system; this has
led to the so-called nonextensive statistical-mechanics formalism
\cite{next:99,next:04,next:05}.  

Among many systems that present unusual behavior, 
one should mention those characterized by anomalous diffusion, e.g., 
particle transport in disordered media.
A possible alternative for describing anomalous-transport processes
consists in introducing modifications in the standard FPE. Within the most
common procedure, one considers nonlinear FPEs \cite{Frank:05}, that in
most of the cases come out as simple
phenomenological generalizations of the usual linear FPE
\cite{Plastino:95,Tsallis:96,Borland:98b,Borland:99,%
LenziMendes:03,Frank:99,Frank:01c,Frank:01b,Malacarne:01,Malacarne:02}. 
In these nonlinear systems, interesting new aspects appear, leading to a 
wide range of open problems, in such a way that their investigation has led
to a new research area in physics, with a lot of applications in natural 
systems.
It is very common to 
find the power-law-like probability distribution that maximizes Tsallis' 
entropy as solutions of some nonlinear FPEs
\cite{Frank:05,Plastino:95,Tsallis:96,Borland:98,Borland:99,LenziMendes:03}. It
seems that the nonextensive statistical mechanics formalism appears to be 
intimately related to nonlinear FPEs, motivating an investigation for a 
better understanding of possible connections between generalized entropies and
nonlinear FPEs~\cite{Compte:96,Frank:00,Frank:01,Martinez:98,Plastino:95,%
Tsallis:96,Frank:99,Frank:01c,Frank:05}. 

Recently, a general nonlinear FPE has been derived directly from a standard
master equation, by introducing nonlinear contributions in the associated
transition probabilities, leading to~\cite{Curado:03,Nobre:04}

\begin{equation}
\frac{\partial P(x,t)}{\partial t} = - \pderiv{(A(x)P(x,t))}{x} + \pderiv{}{x} 
\left( \Omega[P(x,t)] \pderiv{P(x,t)}{x} \right)~; 
\quad \Omega[P]=a \mu P^{\mu-1}+b(2-\nu) P^{\nu-1}~, 
\label{eq:fokk_planck}
\end{equation}

\vskip \baselineskip
\noindent
where $a$ and $b$ are constants, whereas $\mu$ and $\nu$ are real 
exponents 
\footnote{Even though one could use a simpler notation for the
functional $\Omega[P]$, e.g., 
$\Omega[P]=a^{\prime}P^{\mu^{\prime}}+b^{\prime}P^{\nu^{\prime}}$, herein
we will keep the notation of Eq.~(\ref{eq:fokk_planck}), as appeared
naturally in the derivation of the above FPE, for a consistency with
previous publications~\cite{Curado:03,Nobre:04}.}.
The system is in the presence of an external 
potential $\phi(x)$, associated with a dimensionless force
$A(x)=-d\phi(x)/dx$ [$\phi(x)=-\int
\limits_{-\infty}^{x}A(x^{\prime})dx^{\prime}$]; 
herein, we assume analyticity of
the potential $\phi(x)$ and integrability of the force $A(x)$ in all space.
 
In what concerns the functional $\Omega [P(x,t)]$, we are assuming its 
differentiability and integrability with respect to the 
probability distribution $P(x,t)$, in such a way that at least its first
derivative exists, i.e., that it should be at least 
$\Omega[P] \in C^{1}$. Furthermore, this functional should be  
a positive finite quantity, as expected for a proper diffusion-like term;
this property will be verified later on, as 
a direct consequence of the H-theorem.
 
As usual, we assume that the probability
distribution, together with its first derivative, as well as the product
$A(x)P(x,t)$, should all be zero at infinity,   

\begin{equation}
P(x,t)|_{x \rightarrow \pm \infty} = 0~; \quad
\left. {\partial P(x,t) \over \partial x} 
\right|_{x \rightarrow \pm \infty} = 0~; \quad 
A(x)P(x,t)|_{x \rightarrow \pm \infty} = 0~, \quad (\forall t)~. 
\label{eq:p_inf}
\end{equation}

\vskip \baselineskip
\noindent
The conditions above guarantee the preservation of the normalization for the
probability distribution, i.e., if for a given time $t_{0}$ one has
that 
$\int_{-\infty}^{\infty}dx \ P(x,t_{0}) = 1$, then a simple integration of 
Eq.~(\ref{eq:fokk_planck}) with respect to the variable $x$ yields,  

\begin{equation}
\frac{\partial }{\partial t} \int_{-\infty}^{\infty}dx \ P(x,t) = 
- \left[ A(x)P(x,t) \right]_{-\infty}^{\infty} 
+\left( \Omega[P(x,t)] \pderiv{P(x,t)}{x} \right)_{-\infty}^{\infty} = 0~, 
\label{eq:timeder_0}
\end{equation}

\vskip \baselineskip
\noindent
and so, 

\begin{equation}
\int_{-\infty}^{\infty}dx \ P(x,t) = 
\int_{-\infty}^{\infty}dx \ P(x,t_{0}) = 1 \qquad (\forall t)~. 
\label{eq:norm}
\end{equation}

\vskip \baselineskip
\noindent
In the present work we investigate further properties of 
the nonlinear FPE of Eq.~(\ref{eq:fokk_planck}), finding stationary
solutions in several 
particular cases, and discussing its associated entropies, that were
introduced in order to satisfy the H-theorem.  
In the next section we present stationary solutions of this equation; 
in section III we prove the H-theorem by
using Eq.~(\ref{eq:fokk_planck}), and show that the validity of this
theorem can be directly related to the definition of a general entropic form
associated with this nonlinear FPE. In section IV we discuss particular
cases of this general entropic form and their associated nonlinear FPEs.
Finally, in section V we present our conclusions. 

\section{Stationary State}

The nonlinear FPE of Eq.~(\ref{eq:fokk_planck}) is very general and  
covers several particular cases, e.g., the one related to Tsallis' 
thermostatistics~\cite{Plastino:95,Tsallis:96,Borland:98}. 
In this section we will restrict ourselves to a stationary
state, and will derive the corresponding solutions for particular values 
of the parameters associated with this equation.

Let us then rewrite Eq.~(\ref{eq:fokk_planck}) in the form of a continuity
equation,

\begin{equation}
\pderiv{P(x,t)}{t} + \pderiv{j(x,t)}{x} = 0~; \quad 
j(x,t) = A(x) P(x,t) - \Omega[P(x,t)] \pderiv{P(x,t)}{x}~, 
\label{eq:jdef}
\end{equation}

\vskip \baselineskip
\noindent
in such a way that a stationary solution of Eq.~(\ref{eq:fokk_planck}),
$P_{st}(x)$, is associated with a stationary probability flux, 
$j_{st}(x) = {\rm constant}$, which becomes $j_{st}(x) = 0$, when one uses
Eq.~(\ref{eq:p_inf}). Therefore, using the functional $\Omega[P]$ of 
Eq.~(\ref{eq:fokk_planck}), the stationary-state solution
satisfies,

\begin{equation}
A(x) = \left[ a \mu P_{st}^{\mu-2}(x) + b (2-\nu) P_{st}^{\nu-2}(x) \right] 
\pderiv{P_{st}(x)}{x}~,
\end{equation}

\vskip \baselineskip
\noindent
which, after integration, becomes 

\begin{equation}
\phi_0-\phi(x) = a \frac{\mu}{\mu-1} P_{st}^{\mu-1}(x) 
+ b \frac{2-\nu}{\nu-1}P_{st}^{\nu-1}(x)~, 
\label{eq:phi_p}
\end{equation}

\vskip \baselineskip
\noindent
where $\phi_0$ represents a constant. The equation above may be
solved easily in some particular cases, e.g., 
$\nu=\mu$, $\nu=2$, and $\mu=0$,

\begin{equation}
\label{eq:st_sol1}
P_{st}(x) =  \frac{1}{Z^{(1)}} \left[  1 -
      \frac{\phi(x)}{\phi_0}\right]_+^{\frac{1}{\alpha-1}}~; \quad 
Z^{(1)} = \infint dx \left[ 1- \frac{\phi(x)}{\phi_0}
 \right]_{+}^{\frac{1}{\alpha-1}}~, 
\end{equation}

\vskip \baselineskip
\noindent
where $\alpha=\mu$, in the cases $\nu=\mu$ and $\nu=2$, whereas 
$\alpha=\nu$, 
if $\mu=0$. In the equation above, $[ y ]_{+}=y$, for $y>0$, and zero
otherwise. Another type of solution applies for $\mu=2\nu-1$, or
$\nu=2\mu-1$, 
 
\begin{equation}
\label{eq:st_sol2}
P_{st}(x) =   \frac{1}{Z^{(2)}} \left[ 1 \pm
    \sqrt{1+ K \left(
    \phi(x) - \phi_0 \right)} \right]_+^{\frac{1}{\alpha-1}}~; \quad
Z^{(2)} = \infint dx \left[ 1 \pm
    \sqrt{1+ K \left(
    \phi(x) - \phi_0 \right)} \right]_+^{\frac{1}{\alpha-1}}~,
\end{equation}

\vskip \baselineskip
\noindent
where

$$
K = \left\{
\begin{array}{cl}
{{\textstyle 2a} \over {\textstyle b^{2}}} \ 
{{\textstyle (2\nu-1)(\nu-1)} \over {\textstyle (2-\nu)^2}}
\quad , & \ \ {\rm if} \qquad
\mu=2\nu-1 \quad (\alpha=\nu)~,  
\\ & \\
{{\textstyle 2b} \over {\textstyle a^{2}}} \ 
{{\textstyle (3-2\mu)(\mu-1)} \over {\textstyle \mu^2}}
\quad , & \ \ {\rm if} \qquad
\nu=2\mu-1 \quad (\alpha=\mu)~,   
\end{array}
\right.
\eqno(10)
$$

\vskip \baselineskip
\noindent
and we are assuming that $[1+K(\phi(x)-\phi_0)] \geq 0$. 
Some well-known particular cases of the stationary solutions presented
above come out easily,
e.g., from Eq.~(\ref{eq:st_sol1}) one obtains, 
in all three
situations that yielded this equation, the exponential solution associated
with the linear FPE in the limit 
$\alpha \rightarrow 1$ [with $\phi_{0} \propto (\alpha -1)^{-1}$], as well
as the 
generalized exponential solution
related to Tsallis thermostatistics, for $\alpha=2-q$, where $q$ denotes
Tsallis' entropic index. 

\section{H--theorem and the Associated Entropy}

In this section we will demonstrate the H-theorem by making use of 
Eq.~(\ref{eq:fokk_planck}), and for that purpose, an entropic form
related to this equation will be introduced. Let us therefore 
suppose a general entropic form satisfying the following conditions, 

\begin{equation}
S= \infint dx  \ g[P(x)]~;  \,\,\, g[0] = 0~; \,\,\, g[1] = 0~; \,\,\, 
\frac{d^2g}{dP^2} \leq 0~, 
\label{eq:cond_entr}
\end{equation}

\vskip \baselineskip
\noindent
where one should have $g[P(x,t)]$ at least as 
$g[P(x,t)] \in C^{2}$; in addition to that, let us also define the
free-energy functional,  

\begin{equation}
F = U - \frac{1}{ \beta} \ S~; 
\qquad U =  \infint dx \ \phi(x) P(x,t)~,  
\label{eq:free_energy}
\end{equation}

\vskip \baselineskip
\noindent
where $\beta$ represents a Lagrange multiplier, restricted to 
$\beta \geq 0$. Furthermore, we will show
that this free-energy functional is bounded from below; this condition,
together with the H-theorem [$(\partial F/\partial t) \leq 0$), leads, 
after a long time, the system towards a stationary state. 

\subsection{H--Theorem}

The H-theorem for a system that exchanges energy with its surrounding,
herein represented by the potential $\phi(x)$, corresponds to a
well-defined sign for the time derivative of the free-energy
functional defined in Eq.~(\ref{eq:free_energy}). Using the definitions above,

\begin{eqnarray}
\pderiv{F}{t} & = & \pderiv{}{t} \left( \ \infint dx \ 
\phi(x) P(x,t) - \frac{1}
{\beta} \infint dx \  g[P] \right)  \nonumber \\
\nonumber \\
& = & \infint dx \ \left( \phi(x) - \frac{1}{\beta} \pderiv{g[P]}{P} \right) 
\pderiv{P}{t}~.
\end{eqnarray}

\vskip \baselineskip
\noindent
Now, one may use the FPE of Eq.~(\ref{eq:fokk_planck}) for the time
derivative of the probability distribution; 
carrying out an integration by parts, and using the conditions of 
Eq.~(\ref{eq:p_inf}), one obtains, 

\begin{equation}
\pderiv{F}{t} = - \infint dx \ \left[ \deriv{\phi(x)}{x}P(x,t)+\Omega[P]
\pderiv{P}{x} \right]
\left[ \deriv{\phi(x)}{x} - \frac{1}{\beta}  \pderiv{^2g[P]}{P^2} \pderiv{P}{x} \right]~.
\label{eq:HT_step2}
\end{equation}

\vskip \baselineskip
\noindent
Usually, one is interested in verifying the H-theorem from a
well-defined FPE, 
together with a particular entropic form, in such a way that the quantities
$\Omega[P]$ and 
$\partial^{2}g[P]/\partial P^{2}$ are previously defined (see, e.g.,
Refs.~\cite{Frank:05,Shiino:01}). Herein, we follow a different approach, 
by assuming that the general Eqs.~(\ref{eq:fokk_planck}) 
and~(\ref{eq:cond_entr}) should be satisfied; then, we impose the condition,

\begin{equation}
\pderiv{^2g[P]}{P^2} = -\beta \frac{\Omega[P]}{P(x,t)}~, 
\label{eq:rel_entr_omega}
\end{equation}

\vskip \baselineskip
\noindent
in such way that,

\begin{equation}
\pderiv{F}{t} = - \infint dx \ P(x,t) \left[ \deriv{\phi(x)}{x} +
\frac{\Omega[P]}{P(x,t)} \pderiv{P}{x} \right]^{2} \leq 0~.
\label{eq:h_theorem}
\end{equation}

\vskip \baselineskip
\noindent
It should be noticed that Eq.~(\ref{eq:rel_entr_omega}), introduced 
in such a way to provide a well-defined sign for the time derivative of the
free-energy functional, yields two important conditions, as described
next. 

\noindent
(i) $\Omega[P] \ge 0$ 
[cf. Eq.~(\ref{eq:cond_entr})], which is
expected for an appropriate diffusion-like term. 

\noindent
(ii) It expresses a relation involving the FPE of
Eq.~(\ref{eq:fokk_planck}) and an 
associated entropic form, allowing for the calculation of such an entropic
form, given the FPE, and vice-versa. 
Since the FPE is a phenomenological equation that specifies the dynamical
evolution associated with a given physical system, 
Eq.~(\ref{eq:rel_entr_omega}) may be useful in the identification of  
the entropic form associated with such a system.
In particular, the present approach makes it possible to identify 
entropic forms 
associated with some anomalous systems, exhibiting unusual behavior, that
are appropriately described by nonlinear FPEs, like the one of 
Eq.~(\ref{eq:fokk_planck}). 
As an illustration of this point, let us consider  
the simple case of a linear FPE, that describes the dynamical evolution of
many physical systems, essentially those characterized by normal
diffusion. This equation may be recovered     
from Eq.~(\ref{eq:fokk_planck}) by choosing $\mu=\nu=1$, 
in such a way that $\Omega[P]=a+b=D$, where
$D$ represents a positive constant diffusion coefficient with 
units (time)$^{-1}$.
One may now set the Lagrange multiplier $\beta=k_{B}/D$, where $k_{B}$
represents the Boltzmann constant; integrating
Eq.~(\ref{eq:rel_entr_omega}), and using the conditions of 
Eq.~(\ref{eq:cond_entr}), one gets the well-known
BG entropic form, 

\begin{equation}
g[P] = -k_{B} P \ln P~.  
\end{equation}

\vskip \baselineskip
\noindent
In the next section we will explore further the relation of 
Eq.~(\ref{eq:rel_entr_omega}), by analyzing other particular cases. 

The simplest situation for which condition (i) above is satisfied may be
obtained by imposing both terms of the functional $\Omega[P]$ to be
positive, which leads to  

\begin{equation}
a \ge 0~; \ \ \mu \ge 0~, \quad \text{or} \quad a<0~; \ \ \mu < 0~, 
\label{eq:H_cond1}
\end{equation}

\noindent
and 

\begin{equation}
b \ge 0~; \ \ \nu \leq  2~, \quad \text{or} \quad b<0~; \ \ \nu > 2~.  
\label{eq:H_cond2}
\end{equation}

\vskip \baselineskip
\noindent
It should be stressed that it is possible to have $\Omega[P] \ge 0$ with
less restrictive ranges for the parameters above. However, an additional
property
for the free-energy functional of Eq.~(\ref{eq:free_energy}) to be
discussed next, namely, the boundness from below, requires the
conditions of Eqs.~(\ref{eq:H_cond1}) and (\ref{eq:H_cond2}), with the
additional restriction $\nu>0$.  

Now, integrating Eq.~(\ref{eq:rel_entr_omega}) for the general functional
$\Omega[P]$ of Eq.~(\ref{eq:fokk_planck}), and using the
standard conditions for $g[P]$ defined in Eq.~(\ref{eq:cond_entr}), one
gets that 

\begin{equation}
g[P] = - \beta \left[\frac{a}{\mu-1}P^{\mu} + b \frac{2-\nu}
{\nu (\nu-1)}P^{\nu} 
+\frac{a \nu (1-\nu) + b (2-\nu)(1-\mu)}{(1-\mu) (1-\nu) \nu} P \right]~. 
\label{eq:new_entr}
\end{equation}

\vskip \baselineskip
\noindent
This entropic form recovers, as particular cases, the BG entropy (e.g., 
when $\mu, \nu \rightarrow 1$) and 
several generalized entropies defined previously in the literature, 
like those introduced by 
Tsallis~\cite{Tsallis:88}, Abe~\cite{Abe:97}, Borges-Roditi~\cite{Borges:98}, 
and Kaniadakis~\cite{Kaniadakis:01,Kaniadakis:02}. 
Such particular
cases, as well as their associated FPEs, will be discussed in the next section.

For the simpler situation of an isolated system, i.e., 
$\phi(x)=$ constant,  
the H-theorem should be expressed in terms of the time derivative of the 
entropy, in such a way that Eq.~(\ref{eq:HT_step2}) should be replaced by

\begin{equation}
\pderiv{S[P]}{t} = - \int_{-\infty}^{\infty}dx \ \left( \Omega[P] 
\pderiv{P}{x} \right) 
\left( \pderiv{^2 g[P]}{P^2} \pderiv{P}{x} \right)
= - \int_{-\infty}^{\infty}dx \ \Omega[P] \ \pderiv{^2 g[P]}{P^2} 
\left( \pderiv{P}{x} \right)^{2}\geq 0~. 
\end{equation}

\vskip \baselineskip
\noindent
In this case all that
one needs is the standard condition associated with the FPE 
[same condition (i) above], i.e., $\Omega[P] \geq 0$,  
and the general restrictions of Eq.~(\ref{eq:cond_entr}) for the entropy. A
similar result may also be obtained by proving the H-theorem using
the master equation from which  
Eq.~(\ref{eq:fokk_planck}) was derived, with the transition probabilities
introduced in  
Ref.~\cite{Curado:03,Nobre:04} (see the Appendix). 

\subsection{Boundness from Below}

Above, we have proven that the free-energy functional decreases in time,
and so, for the existence of a stationary state at long times of an
evolution process, characterized by a probability distribution $P_{st}(x)$,
one should have that

\begin{equation}
F(P(x,t)) \geq F(P_{st}(x)) \qquad (\forall t). 
\label{eq:bound_freeenergy}
\end{equation}

\vskip \baselineskip
\noindent
In what follows, we will show this inequality and find the conditions for
its validity. 
Therefore, using Eqs.~(\ref{eq:phi_p}), (\ref{eq:cond_entr}), and 
(\ref{eq:free_energy}), we can write,

\begin{equation}
F(P(x,t)) = \infint P(x,t) \left( \phi_0 - a\frac{\mu}{\mu-1} 
P_{st}^{\mu-1}(x) 
- b \frac{2-\nu}{\nu-1} P_{st}^{\nu-1}(x) \right) dx 
- \frac{1}{\beta} \infint g[P] dx~,
\end{equation}

\vskip \baselineskip
\noindent
and so,

\begin{equation}
F(P_{st})-F(P)= \infint \left( P - P_{st} \right) \left( a
  \frac{\mu}{\mu-1} P_{st}^{\mu-1} + b \frac{2-\nu}{\nu-1} P_{st}^{\nu-1}
\right) dx + \frac{1}{\beta} \infint \left( g[P]-g[P_{st}] \right) dx~,
\end{equation}

\vskip \baselineskip
\noindent
where we have used the normalization condition for the probabilities.
Now, we insert the entropic form of Eq.~(\ref{eq:new_entr}) in the equation
above to obtain,

\begin{equation}
F(P_{st})- F(P) = \infint \left[ \frac{a}{\mu-1} P_{st}^{\mu} \,
  \Gamma_{\mu}[P/P_{st}] + \frac{b (2-\nu)}{\nu (\nu-1)} P_{st}^{\nu} 
  \, \Gamma_{\nu}[P/P_{st}]
  \right] dx~,
\label{eq:bound_end}
\end{equation}

\vskip \baselineskip
\noindent
where,
\begin{equation}
\Gamma_{\alpha}[z]= 1- \alpha+ \alpha z - z^{\alpha} \qquad (\alpha=\mu,\nu)~.
\label{eq:alpha_eq}
\end{equation}

\vskip \baselineskip
\noindent
By analyzing the extrema of the functional $\Gamma_{\alpha}[P/P_{st}]$, one
can see that $\Gamma_{\alpha}[P/P_{st}] \geq 0$ for 
$0 < \alpha < 1$ and $\Gamma_{\alpha}[P/P_{st}] \leq 0$ for 
$\alpha > 1$ and $\alpha < 0$. Therefore, the inequality of
Eq.~(\ref{eq:bound_freeenergy}) is satisfied for the range of parameters 
specified by Eqs.~(\ref{eq:H_cond1}) and (\ref{eq:H_cond2}), if one
considers additionally, $\nu > 0$.  

For the case of an isolated system, the stationary solution turns out to be
the equilibrium state, which is the one that maximizes the entropy.
Therefore, one may use the concavity property 
of the entropy [cf. Eq.~(\ref{eq:cond_entr})] in order to get, 

\begin{equation}
S[P_{eq}(x)]-S[P(x,t)] = \infint \left( g[P_{eq}]-g[P] \right) dx \ge 0
\qquad (\forall t). 
\label{eq:bound_entropy}
\end{equation}

\section{Some Particular Cases}

In this section we analyze some particular cases of the entropic form of 
Eq.~(\ref{eq:new_entr}) and using Eq.~(\ref{eq:rel_entr_omega}), we 
find for each of them, the corresponding functional
$\Omega[P]$ of the associated FPE. In the examples that follow, we will set
the Lagrange multiplier $\beta=k/D$, with $k$ and $D$
representing, respectively, a constant with dimensions of entropy and 
a constant diffusion coefficient. 

\vskip \baselineskip
\noindent
(a) Tsallis entropy~\cite{Tsallis:88}: This represents the most
well-known generalization of the BG entropy, which
has led to the development of the area of nonextensive statistical
mechanics \cite{next:99,next:04,next:05}. 
One may find easily that Eq.~(\ref{eq:new_entr}) recovers Tsallis
entropy in several particular cases, e.g., 
$\{b=0,a=D,\mu=q\}$, $\{a=0,b=D\nu /(2-\nu),\nu=q \}$, 
and $\{a=D/2, b=D\nu/[2(2-\nu)], \mu=\nu=q \}$. For all these cases one may
use Eq.~(\ref{eq:rel_entr_omega}), in order to get the corresponding 
functional $\Omega[P]$,
 
\begin{equation}
g[P] = k \frac{P^q- P}{1-q} , \,\,\,\ \Omega[P] = q D P^{q-1}~.   
\end{equation}

\vskip \baselineskip
\noindent 
With the functional $\Omega[P]$ above, one identifies the nonlinear FPE
that presents the well-known $q$-exponential, or Tsallis
distribution (replacing $q \rightarrow 2-q$), as a time-dependent solution 
\cite{Plastino:95,Tsallis:96,Frank:99}. 

\vskip \baselineskip
\noindent
(b) Abe entropy~\cite{Abe:97}: This proposal was inspired in the area of 
quantum groups, where certain quantities, usually called $q$-deformed
quantities, are submitted to deformations and are often
required  
to possess the invariance $q \leftrightarrow q^{-1}$.  
The Abe entropy may be obtained from Eq.~(\ref{eq:new_entr}) in the 
particular case $\{a=D(q-1)/(q-q^{-1}), b=-Dq(q+1)/[(q-q^{-1})(q+2)],
\mu=-\nu=q\}$, for which 

\begin{equation}
g[P]= - k \frac{P^q-P^{-q}}{q-q^{-1}}, \,\,\, \Omega[P]= D \left(
  \frac{q(q-1)}{q-q^{-1}} P^{q-1} - \frac{q (q+1)}{q-q^{-1}} P^{-q-1} 
  \right)~.  
\end{equation}

\vskip \baselineskip
\noindent
(c) Borges-Roditi entropy~\cite{Borges:98}: This consists in another
generalization of Tsallis entropy, where now one has two distinct entropic
indices, $q$ and  
$q^{\prime}$, with a more general invariance $q \leftrightarrow q^{\prime}$; 
this case may be obtained from Eq.~(\ref{eq:new_entr}) by choosing 
$\{a=D(q-1)/(q^\prime-q), b=Dq^\prime (q^\prime -1)/[(q-q^\prime) (2-q^\prime)], \mu=q, \nu=q^\prime \}$. One gets, 

\begin{equation}
g[P] = - k \frac{P^q-P^{q^\prime}}{q-q^\prime}, \,\,\,
\Omega[P]=D \frac{1}{q-q^\prime} \left( q(q-1) P^{q-1} -
q^\prime(q^\prime-1)P^{q^\prime-1} \right)~.  
\end{equation}

\vskip \baselineskip
\noindent
(d) Kaniadakis entropy~\cite{Kaniadakis:01,Kaniadakis:02}: This is also a
two-exponent entropic form, but slightly different from those presented in
examples (b) and (c) above;  
it may be reproduced from Eq.~(\ref{eq:new_entr}) by choosing 
$\{a=b=D/[2(1+q)], \mu=1+q, \nu=1-q\}$, in such a way that

\begin{equation}
g[P] = - \frac{k}{2 q} \left( \frac{1}{1+q}
  P^{1+q} - \frac{1}{1-q} P^{1-q} \right), \,\,\, \Omega[P] =
\frac{D}{2} (P^{q} + P^{-q})~. 
\end{equation}

\vskip \baselineskip
Except for the well-known example (a), i.e., Tsallis entropy and its
corresponding FPE, the other three particular cases presented herein were
much less explored in the literature. Their associated FPEs, defined in
terms of their respective functionals  
$\Omega[P]$ above are, to our knowledge, presented herein for the first
time. These equations, whose nonlinear terms depend essentially in two
different powers of the probability distribution, may be appropriated for
anomalous-diffusion phenomena where a crossover between two different
diffusion regimes occurs~\cite{LenziMendes:03}.  

\section{Conclusion}

We have analyzed important aspects associated with a recently introduced
nonlinear Fokker-Planck equation, that was derived 
directly from a master equation by setting nonlinear effects on its
transition rates.  Such equation is characterized
by a nonlinear diffusion term that  
may present two distinct powers of the probability
distribution; for this reason, it may reproduce, as particular cases, a large
range of nonlinear FPEs of the literature.  
We have obtained stationary solutions for this equation in several cases,
and some of them recover the well-known Tsallis distribution. 
We have proven the H-theorem, and for that, an important relation involving
the parameters of the FPE and an entropic form was introduced. 
Since the FPE is a phenomenological equation that specifies the dynamical
evolution associated with a given physical system, such a relation 
may be useful for identifying entropic forms associated with real
systems and, in particular, anomalous systems that exhibit unusual
behavior and are appropriately described by nonlinear FPEs. 
It is shown that, in the simple case of a linear FPE, the Boltzmann-Gibbs
entropy comes out straightforwardly from this relation.  
Considering the above-mentioned nonlinear diffusion term, this relation
yields a very general entropic form, which similarly to its corresponding
FPE, depends on two distinct powers of the probability distribution. Apart
from Tsallis' entropy, other entropic forms introduced in the literature are
recovered as particular cases of the present one, essentially those
characterized by two entropic indices. Nonlinear FPEs (as well as their
associated entropic forms) whose nonlinear terms depend essentially on two
different powers of the probability distribution, like the ones discussed
in the present paper, are good candidates for describing
anomalous-diffusion phenomena where a crossover between two different
diffusion regimes may take place. As a typical example, one could have a
particle transport in a system composed by two types of disordered media,
characterized respectively, by significantly different grains (e.g.,
different average sizes), arranged in such a way that the diffusion process
is dominated by one medium, at high densities, and by the other one,
at low densities.  

\bibliographystyle{unsrt}
\bibliography{entropy2}

\section*{Appendix}

In this appendix we will prove the H-theorem directly from the master
equation, for an isolated system (i.e., no external forces). Let us
consider  
a system described in terms of $W$ discrete stochastic variables; then,  
$P_{n}(t)$ represents the probability of finding this system in a state
characterized by the  
variable $n$ at time $t$. This probability evolves in time according to a
master equation,   

\begin{equation}
\frac{\partial P_n}{\partial t} = \sum \limits_{m=1}^W \left( P_m w_{m,n} - P_n w_{n,m} \right)~,
\end{equation}

\vskip \baselineskip
\noindent
where $w_{k,l}(t)$ represents the transition probability rate from state $k$ to 
state $l$. The nonlinear effects were introduced through the transition 
rates~\cite{Curado:03}, 

\begin{equation}
w_{k,l}(\Delta) = {1 \over \Delta ^{2}} (\delta _{k,l+1} + \delta _{k,l-1})
[aP^{\mu -1}_{k}(t) + bP^{\nu -1}_{l}(t)]~,  
\end{equation}

\vskip \baselineskip
\noindent
where $\Delta$ represents the size of the step of the random walk. Herein, 
we shall consider a random walk characterized by
$\Delta=1$; substituting such a transition rate in the master equation one
gets,   

\begin{align}
n=1: \ \ & \frac{\partial P_1}{\partial t} = a P_2^{\mu} - a P_1^{\mu} + b P_2
P_1^{\nu-1} - b P_1 P_2^{\nu-1}~,   
\label{eq:discr_master1}\\
n=W: \ \ & \frac{\partial P_W}{\partial t} = a P_{W-1}^{\mu} - a P_W^{\mu} + b P_{W-1}
P_W^{\nu-1} - b P_W P_{W-1}^{\nu-1}~,   
\label{eq:discr_master2}\\
n=2, ..., (W-1): \ \ & \frac{\partial P_n}{\partial t} =  a
\left( P_{n+1}^{\mu} + P_{n-1}^{\mu} \right) - 2a P_n^{\mu} + b P_n^{\nu-1}
\left( P_{n+1} + P_{n-1} \right) \nonumber \\ 
& - b P_n \left( P_{n+1}^{\nu-1}
  + P_{n-1}^{\nu-1} \right)~, 
\label{eq:discr_master3}
\end{align}

\vskip \baselineskip
\noindent
where we have treated the borders of the spectrum ($n=1$ and $n=W$)
separately from the rest. 
Let us now consider the entropy, $S= \sum \limits_{n=1}^{W} g[P_n]$, satisfying the same
conditions specified in the text [see Eq.~(\ref{eq:cond_entr})]; the H-theorem, to be proven below, is expressed in terms of its time derivative, 

\begin{equation}
\pderiv{S}{t} = \sum \limits_{n=1}^{W} \frac{dg[P_{n}]}{dP_{n}} 
\frac{d P_{n}}{d  t} \ge 0~.  
\label{eq:tderiv_entropy}
\end{equation}

\vskip \baselineskip
\noindent
Then, using Eqs.~(\ref{eq:discr_master1})--(\ref{eq:discr_master3}) one can
write this time derivative as, 

\begin{equation}
\begin{aligned}
\pderiv{S}{t} = \sum \limits_{n=1}^{W-1} \frac{dg[P_n]}{dP_n} 
\left( a P_{n+1}^{\mu} - a
  P_n^{\mu}   - b P_n P_{n+1}^{\nu-1}
+ b P_{n+1}  P_n^{\nu-1} \right) \\
+  \sum \limits_{n=2}^{W} \frac{dg[P_n]}{dP_n} \left(
a P_{n-1}^{\mu} - a  P_n^{\mu}  + b P_{n-1} P_n^{\nu-1} 
- b  P_n  P_{n-1}^{\nu-1}\right)~. 
\end{aligned}
\end{equation}

\vskip \baselineskip
\noindent
The sum indices can be rearranged to yield all summations in the range 
$\sum \limits_{n=1}^{W-1}$, 
and thus the time derivative of the entropy becomes,

\begin{equation}
  \pderiv{S}{t} = \sum \limits_{n=1}^{W-1} \left( \frac{dg[P_n]}{dP_n} -
    \frac{dg[P_{n+1}]}{dP_{n+1}} \right) \left[ a \left( P_{n+1}^{\mu} -
      P_{n}^{\mu} \right)  - b P_n P_{n+1} \left(
P_{n+1}^{\nu-2} -P_{n}^{\nu-2}   \right)   \right]~.
\label{eq:brackets_entropy}
\end{equation}

\vskip \baselineskip
\noindent
The negative curvature of the entropic function 
[cf. Eq.~(\ref{eq:cond_entr})] implies that its first derivative decays
monotonically with $P_n$. Hence, for $P_{n+1} > P_n$, the condition of 
Eq.~(\ref{eq:tderiv_entropy}) is satisfied for

\begin{equation}
a \frac{ P_{n+1}^{\mu} - P_{n}^{\mu}}{P_{n+1}-P_n}  - b P_n P_{n+1} \frac{P_{n+1}^{\nu-2}
  -P_{n}^{\nu-2} }{P_{n+1}-P_n} \geq 0 ~, 
\end{equation}

\vskip \baselineskip
\noindent
where we divided the term inside the brackets in Eq.(\ref{eq:brackets_entropy}) by 
the difference $\Delta P =P_{n+1}-P_n$. Now we consider the
limit $\Delta P \rightarrow 0$ and obtain,

\begin{equation}
a \mu P^{\mu-1} + b (2-\nu) P^{\nu-1} \geq 0~, 
\end{equation}

\vskip \baselineskip
\noindent
which corresponds to the condition $\Omega[P] \ge 0$ found in the text, when proving the H-theorem by making use of the FPE of Eq.~(\ref{eq:fokk_planck}).
It should be mentioned that the procedure above works also for $P_{n} > P_{n+1}$~; therefore, in this appendix we have proven that the H-theorem holds for an arbitrary state $n$ and all times $t$ of an isolated system.

\end{document}